\documentclass[aps,groupedaddress,superscriptaddress,amssymb,amsmath,amsbsy,amsfonts,twocolumn,pra]{revtex4}

\usepackage{amssymb,amsmath,amsthm,color,graphicx,times,graphicx}
\usepackage{hyperref}
\usepackage{enumerate}
\usepackage{subfigure}
\usepackage{dsfont}
\usepackage{times}
\usepackage{txfonts}
\usepackage{bbold}

\usepackage{extarrows}

\newcommand{\ket}[1]{\left|#1\right>}
\newcommand{\bra}[1]{\left<#1\right|}

\newcommand*{\tn}[1]{{\textnormal{#1}}}

\theoremstyle{plain}

\theoremstyle{definition}

\date{\today}

\begin{document}

\title{Entropic comparison of Landau--Zener and Demkov interactions in the phase space of a quadrupole billiard}

\author{Kyu-Won Park}
\affiliation{Research Institute of Mathematics, Seoul National University, Seoul 08826, Korea}
\author{Juman Kim}
\affiliation{Department of Physics and Astronomy \& Institute of Applied Physics, Seoul National University, Seoul 08826, Korea}
\author{Jisung Seo}
\affiliation{Department of Physics and Astronomy \& Institute of Applied Physics, Seoul National University, Seoul 08826, Korea}
\author{Songky Moon}
\affiliation{Faculty of Liberal Education, Seoul National University, Seoul 08826, Korea}
\author{Kabgyun Jeong}
\email{kgjeong6@snu.ac.kr}
\affiliation{Research Institute of Mathematics, Seoul National University, Seoul 08826, Korea}
\affiliation{School of Computational Sciences, Korea Institute for Advanced Study, Seoul 02455, Korea}
\author{Kyungwon An}
\affiliation{Department of Physics and Astronomy \& Institute of Applied Physics, Seoul National University, Seoul 08826, Korea}

\date{\today}

\begin{abstract}
We investigate two types of avoided crossings in a chaotic billiard within the framework of information theory. The Shannon entropy in the phase space for the Landau--Zener interaction increases as the center of the avoided crossing is approached. Meanwhile, that for the Demkov interaction decreases as the center of avoided crossing is passed by with an increase in the deformation parameter. This feature can provide a new indicator for scar formation. In addition, it is found that the Fisher information of the Landau--Zener interaction is significantly larger than that of the Demkov interaction.
\end{abstract}

\maketitle

\section{\label{sec:level1}Introduction}
Avoided crossing (AC) is a phenomenon where the two eigenvalues of a Hamiltonian on a specific system parameter come closer but repel each other as the parameter is varied~\cite{JE29}. Avoided crossing is a fundamental and important notion in quantum mechanics because it manifests the existence of an interaction or perturbation between states in a physical system. It has been broadly investigated theoretically and experimentally in numerous systems such as crystalline solids~\cite{MB19}, dielectric gratings~\cite{EM18}, and plasmonic-photonic systems~\cite{SV18}.

There are mainly two types of avoided crossings: the Landau--Zener avoided crossing and Demkov avoided crossing~\cite{YV68}. The familiar Landau--Zener avoided crossing occurs over a narrow parameter range between two eigenfunctions that exchange their features~\cite{JE29,FM17}. On the other hand, the Demkov avoided crossing occurs between two eigenfunctions over a long parameter range, thereby creating a new pair of localized eigenfunctions on periodic orbits~\cite{YV68,FF97,YH15}, \textit{i.e.,} a stable periodic orbit and unstable periodic orbit. Thus, this avoided crossing can support the formation of a scar in quantum chaos~\cite{FF97,FF98}, \textit{i.e.,} wavefunction localization on unstable periodic orbits~\cite{E84}. However, the best of our knowledge, no quantitative or elaborate comparison between these two avoided crossings has been performed. Hence, it is worthwhile to study these simultaneously to better understand their fundamental characteristics.

\begin{figure*}
\centering
\includegraphics[width=13.0cm]{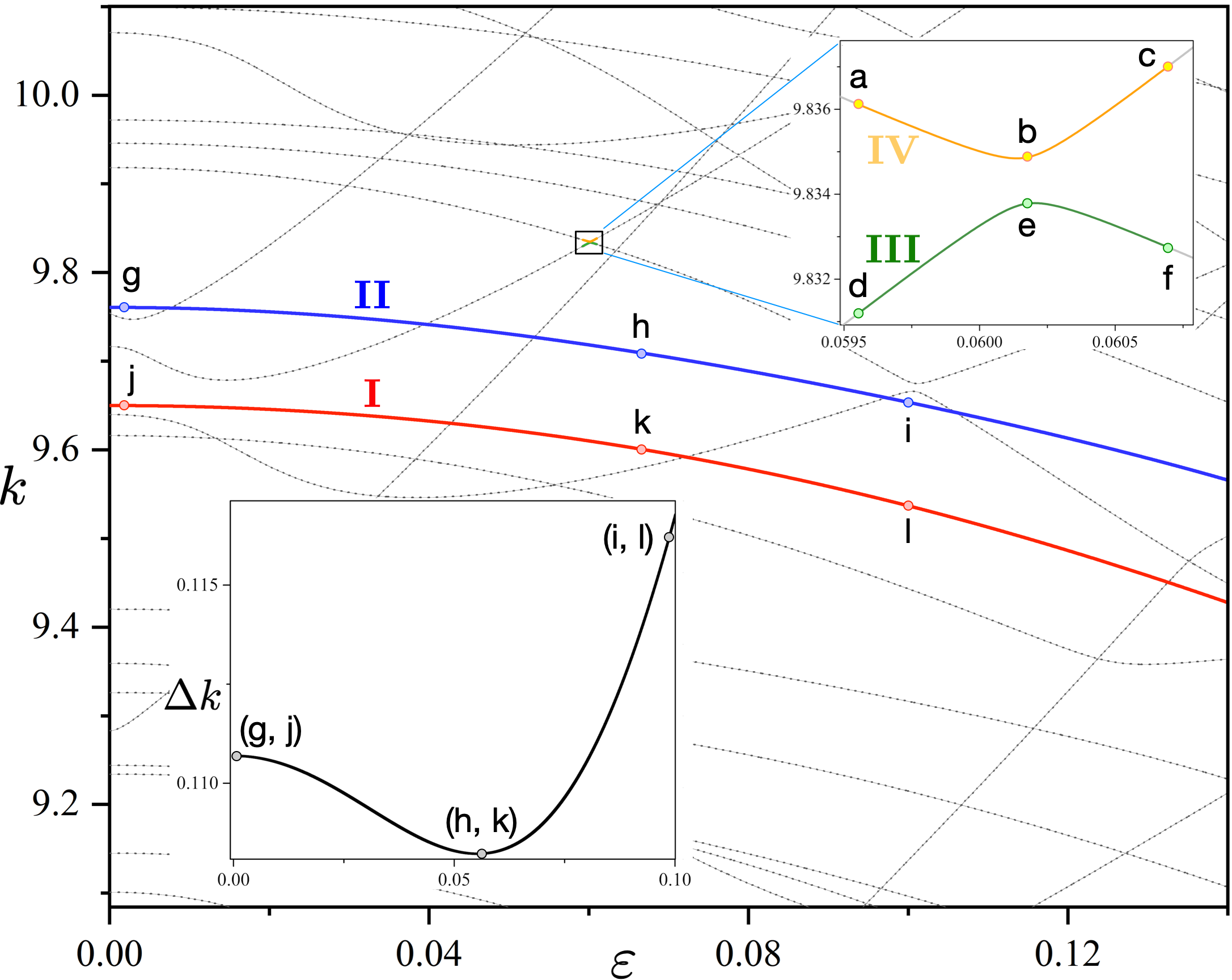}
\caption{Eigenvalue trajectories depend on the deformation parameter $\epsilon$ in a quadrupole billiard. The red solid curve \uppercase\expandafter{\romannumeral1} and blue solid curve \uppercase\expandafter{\romannumeral2} are a pair of avoided crossings for Demkov interactions. The lower inset panel shows the relative differences between \uppercase\expandafter{\romannumeral1} and \uppercase\expandafter{\romannumeral2}. The upper inset panel is an extension of the eigenvalue trajectories corresponding to the Landau--Zener interactions.}
\label{Figure-1}
\end{figure*}

An avoided crossing is defined in terms of eigenvalues. However, we can intuitively consider that the corresponding eigenfunctions also undergo certain alterations during the avoided crossing. In our previous papers, we studied the Shannon entropies related to the avoided crossing~\cite{KS18}, exceptional point~\cite{KJ20}, and wave-function localization~\cite{KJ21}. These works have been investigated only in the configuration space. However, in order to properly describe the dynamical properties of quantum billiards including scar formations~\cite{YH15,FF98,E84}, Fermi resonance~\cite{YH15,JJ17}, and resonance-assisted-tunneling (RAT)~\cite{OP01,OP02,KY15}, the configuration space is not appropriate. Thus, we must consider a phase space for this analysis.

In this study, we extend the investigation to phase space using Husimi distributions~\cite{K40,BG93,AS04} and utilize Shannon entropy to analyze the two types of avoided crossings, studying the relation between localization properties and overlaps of Husimi distributions. It is observed that unlike the case of Landau--Zener avoided crossing the Shannon entropies of a pair of Demkov avoided crossing decrease as the center of the avoided crossing is passed by. Furthermore, the Shannon entropies of the unstable periodic orbit are smaller than that of the stable periodic orbit throughout the procedure, without exchange of Shannon entropies. Thus, the Shannon
entropy could be used as a new indicator of scar formation. Furthermore, we quantitatively describe the differences between narrow-range interactions and long-range interactions by measuring the sensitivity of normalized Husimi distributions to variations in their underlying parameters. For this approach, we introduce the Fisher information~\cite{R25}.

The remainder of this paper is organized as follows. In Sec.~II, we briefly introduce the feature of the Demkov interaction. In Sec.~III, two types of avoided crossings in a quadrupole billiard are addressed. We study the overlap of eigenfunctions and that of Husimi distributions in Sec.~IV. The Shannon entropies in the phase space are discussed in Sec.~V. In Sec.~VI, we discuss the Fisher information of Demkov and Landau--Zener interactions. Finally, we summarize our works in Sec.~VII.

\section{A brief introduction to Demkov interaction for dynamical system}
\subsection{Recapitulation of Demkov interaction compared with Landau--Zener interaction}

At first, let us briefly recapitulate the feature of the Demkov interaction, comparing with the well-know Landau--Zener interaction. The formula associated with Landau--Zener interaction is a solution governing the transition dynamics of a two-level quantum system~\cite{JE29}. Thus, the eigenvalue trajectories of the Landau--Zener interaction are just solutions of two by two Hamiltonian ($\textsf{H}_{LZ}$). For the simplification, the difference between two eigenvalues are linearly proportional to the time: $\Delta E=E_{1}-E_{2}=\xi t$, and under this condition, the probability of a diabatic transition is given by $T_{LZ}=\exp^{-2\pi\eta}$ with $\eta=\frac{v^{2}}{\hbar|\xi|}$~\cite{JE29}. Here, $v$ is an off-diagonal term of $\textsf{H}_{LZ}$. It should be noticed that the two-level quantum system in a real physical system can be justified only when the magnitude of the level repulsion is so small compared with the distance to the other levels.

On the other hand, the Demkov interaction can take places in the opposite case to the Landau--Zener interaction, i.e., the magnitude of the level repulsion is much larger compared with the distance to the other levels so that the overall shapes of interacting two modes look so broad~\cite{YV68}. Moreover, this condition indicates that the Demkov interaction is generally not restricted to two-level systems, yielding non-isolated avoided crossings~\cite{FF98,JJ17} because the ending part of one of the interacting two level can interact with other levels in general. The perturbed term of Hamiltonian for Demkov interaction ($\textsf{H}_{D}$) is given by $\beta t\ket{\varphi}\bra{\varphi}$ with $\beta>0$ and $t$ as a parameter. Here, $\ket{\varphi}\bra{\varphi}$ is a projection operator onto a zero state (initial state) $\ket{\varphi}$. Hence, the eigenvalue equation of $\textsf{H}_{D}$ is given by $[(\textsf{H}_{0}-E)+\beta t\ket{\varphi}\bra{\varphi}]\ket{\psi}=0$ with a non-perturbed Hamiltonian $\textsf{H}_{0}$. Accordingly, the eigenvalues are poles of the function $\bra{\varphi}G(E)\ket{\varphi}$ on a contour integration, where the $G(E)=(\textsf{H}_{0}-E)^{-1}$ is a resolvent operator. See the details in the reference~\cite{YV68}. Furthermore, the transition probability from the zero state $\ket{\varphi}$ to another state $\ket{n}$ with an energy larger than $E$ is given by $T_{D}=\exp(-2\pi\beta^{-1}\sum_{E_{n}<E}\textsf{H}_{n}^{2})$ with $\textsf{H}_{n}=\bra{n}\textsf{H}_{D}\ket{0}$~\cite{YV68}. Thus, $T_{D}$ goes to $1$ as $\beta$ goes to infinity.

\subsection{Fermi resonance and RAT as implementing Demkov interaction in quantum billiard}

Implementing Demkov interaction in a quantum billiard system is not addressed well mathematically or quantitatively up to now. Instead of that, indirectly, Fermi resonance~\cite{YH15,JJ17} and RAT~\cite{KY15,FR19} have been studied in a quantum billiard under the Demkov interaction. Thus, we exploit the Fermi resonance and RAT to investigate the features of the Demkov interaction in this paper.

The avoided crossings associated with classical resonance in the phase space can be formulated in terms of Fermi resonance. The Hamiltonians of two eigenstates can be described by $H(I_{1}, I_{2})$ and $H(I'_{1}, I'_{2})$ where the $I_{i}$ and $I'_{i}$ are action variables for the two eigenstates. Hence, the crossing of a pair of eigenstates indicates semiclassically $H(I_{1}, I_{2})=H(I'_{1}, I'_{2})$. When $|I_{i}-I'_{i}|$ is small, we can expand $H(I'_{1}, I'_{2})$ around $I'_{i}=I_{i}$ to yield: $H(I'_{1}, I'_{2})\simeq H(I_{1}, I_{2})+(I_{1}-I'_{1})\omega_{1}+(I_{2}-I'_{2})\omega_{2}+$...,
where $\omega_{i}=\partial H/\partial I_{i}$ are the frequencies associated with the actions $I_{i}$. Substituting the semiclassical quantization condition $H_{i}=(n_{i}+\frac{\alpha_{i}}{4})$ with Maslov index $\alpha_{i}$ result in the relation: $(n_{1}-n'_{1})\omega=(n_{2}-n'_{2})$~\cite{YH15,JJ17}. If the winding number $\omega_{1}/\omega_{2}$ is rational, the corresponding orbit is periodic, and it gives rise to the stable orbit or unstable orbit (i.e., scar).

The RAT induces an enhanced interaction of unperturbed modes (WGMs) that live along nearby invariant tori, owing to nonlinear resonance chains located between unperturbed modes when the quantum number of unperturbed modes differ in a multiple of the order of the resonance chain (selection rule)~\cite{OP01,OP02,KY15}, and the selection rules are associated with Fermi resonance~\cite{FF98,CY17}. Hence, RAT also implies a leading to the wavefunction localization on the periodic orbits associated with the nonlinear resonance chain~\cite{DM11,D14}.

It should be stressed that because both of the Fermi resonance and the RAT need well-defined classical resonance chains or invariant tori, and also quantum numbers, the Demkov type interactions take place in a regular or mixed phase space rather than a fully chaotic phase space.

\section{Two types of interactions in a quantum billiard}

\subsection{Introduction to the dynamical billiards}
Let us briefly introduce a dynamical billiard in a classical and quantum system, respectively, before addressing the interactions in the quantum billiard in detail. A dynamical billiard is a dynamical system in which particles move through without loss of energy as a straight line. Moreover, when the particles hit the boundary, they reflect from it without loss of energy and all reflections are specular, i.e., the incident angle just before the collision is equal to the reflected angle just after the collision. The classical Hamiltonian $H$ for describing a particle of mass $m$ moving freely without friction in billiard system ($\Omega$) is $H(q,p)=\frac{p^{2}}{2m}+v(q)$ where $v(q)$ is zero inside the billiards and infinity otherwise: $v(q)=0$ when $q\in \Omega$ and $v(q)=\infty$ when $q \notin\Omega$. This infinity potential ensures a specular reflection on the boundary of the billiard system.

For a dynamical billiard in quantum systems, the Hamiltonian equations is replaced by the time independent Schr\"{o}dinger equation: $-\frac{\hbar^{2}}{2m}\nabla^{2}\psi_{n}(q)=E_{n}\psi_{n}(q)$. The potential $v(q)$ given above lead to the Dirichlet boundary conditions: $\psi_{n}(q)=0$ for all $q \notin\Omega$ and also to the Helmholtz equation $(\nabla^{2}+k^{2})\psi=0$ with a wave number $k=\frac{1}{\hbar}\sqrt{2mE_{n}}$.

If the classical billiards belong to integrable systems such as a circular and rectangular system, then the corresponding quantum billiards are completely solvable and the spectral properties coincide with those of Poissonian random numbers. On the other hand, when the classical billiard is chaotic, then the corresponding quantum billiards are generally not exactly solvable and the spectral properties coincide with those of random matrices from the Gaussian ensembles~\cite{F10}.

\begin{figure}
\centering
\includegraphics[width=8.5cm]{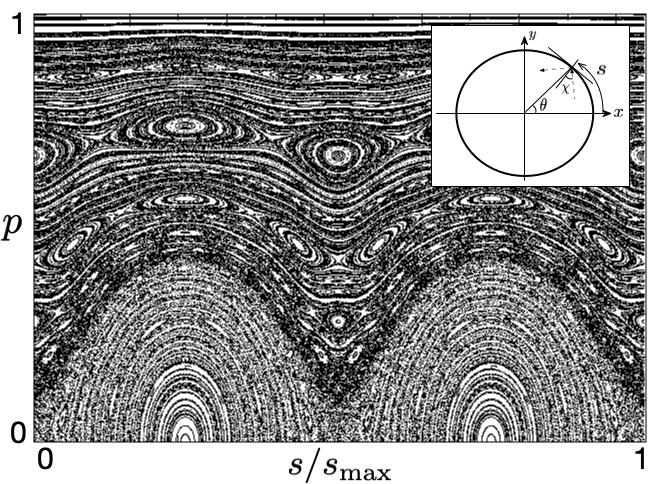}
\caption {An example of Poincar\'{e}-surface of section (classical phase space) defined in terms of Poincar\'{e}-Birkhoff coordinates at $\varepsilon=0.05$. The inset is a schematic diagram for Poincar\'{e}-Birkhoff coordinate: $s$ is a boundary length of the cavity from the $x$-axis and its conjugate momentum $p = \sin(\chi)$. Here, $\chi$ is the incident angle with respect to the normal to the boundary wall, and $s_{\max}$ is the total length of the cavity boundary.}
\label{Figure-2}
\end{figure}

\subsection{Eigenvalues and eigenfunctions of a chaotic billiard}
In this study, we considered a closed quadrupole cavity as a model for a chaotic quantum billiard in a Hermitian system, because this quadrupole billiard has been extensively exploited for studying the physical phenomena in the phase space~\cite{PN03,WLL+21} and for the candidate of deformed micro cavity lasers~\cite{ZYX+09,CW15}. Its geometrical boundary shape is described by $\rho(\theta)=(1+\epsilon\cos(2\theta))$. Here, $\theta$ is an angle in polar coordinates and $\epsilon$ is the deformation parameter. We numerically calculated the eigenvalues and their eigenfunctions using the boundary element method~\cite{J03} by solving the Helmholtz equation $\nabla^{2}\psi+n^{2}k^{2}\psi=0$ for transverse-magnetic modes that satisfy the Dirichlet boundary conditions. Here, $\psi$ is the vertical component of the electric field, $k$ denotes the vacuum wavenumber, and $n=3.3$ is the refractive index of the cavity. The eigenvalue trajectories in the region $k\in [9.1, 10.1]$ are plotted as a function of the deformation parameter $\epsilon$ in Fig.~\ref{Figure-1}.

The red solid curve \uppercase\expandafter{\romannumeral1} and blue solid curve \uppercase\expandafter{\romannumeral2} are a pair of eigenvalue trajectories for the Demkov interactions. We cannot easily identify avoided crossing between these two curves. However, the black solid curve in the lower inset panel in Fig.~\ref{Figure-1} resolves this problem. It is the relative difference between \uppercase\expandafter{\romannumeral1} and \uppercase\expandafter{\romannumeral2}, and is minimized at $\epsilon\simeq0.055$. This indicates that the coupling strength of the Demkov interactions is maximized at $\epsilon\simeq0.055$~\cite{JJ17}. The green solid curve \uppercase\expandafter{\romannumeral3} and orange solid curve \uppercase\expandafter{\romannumeral4} in the upper inset panel are a pair of eigenvalue trajectories for the Landau--Zener interactions. We observe that the Landau--Zener interaction occurs in a highly narrow parameter region compared with the Demkov interaction. Note that the center of the avoided crossing is located at $\epsilon\simeq0.0603$.

\begin{figure*}
\centering
\includegraphics[width=18cm]{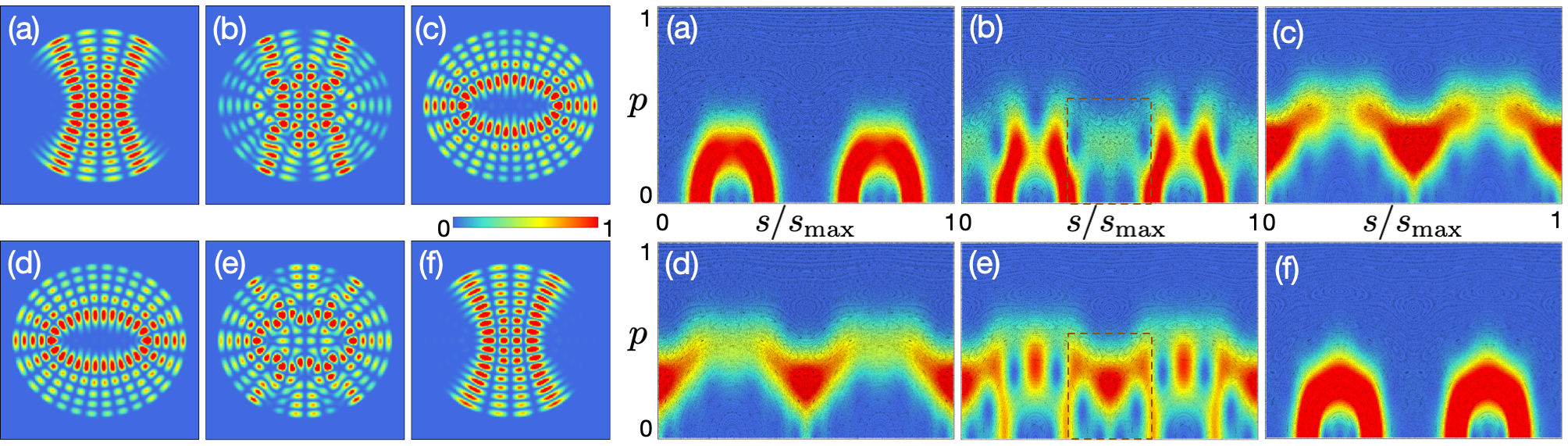}
\caption {Intensities of certain representative eigenfunctions and their Husimi distributions for the Landau--Zener interaction. The panels (a), (b), (c), (d), (e), and (f) on the left side are the intensities of certain representative eigenfunctions in Fig.~\ref{Figure-1}. The panels (a), (b), (c), (d), (e), and (f) on the right side are Husimi distributions corresponding to each intensity of the eigenfunctions in the left panel. The classical phase spaces (Poincar\'{e}-surface of section) are displayed  that are superimposed with Husimi distributions at right panels.}
\label{Figure-3}
\end{figure*}

\begin{figure*}
\centering
\includegraphics[width=18cm]{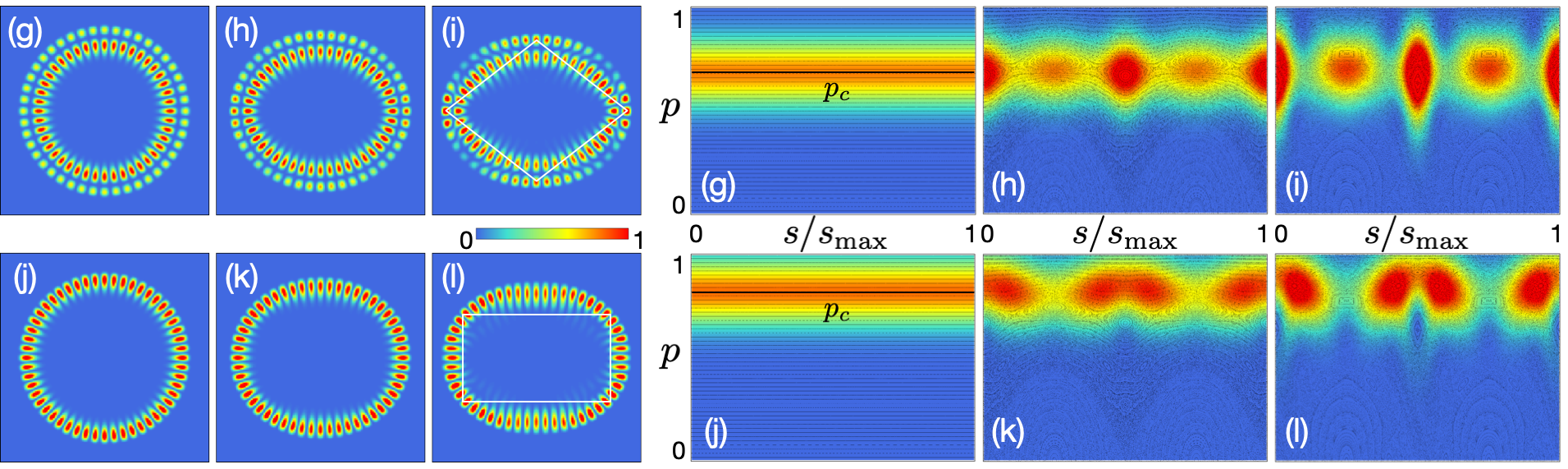}
\caption {Intensities of certain representative eigenfunctions and their Husimi distributions for the Demkov interaction. The left panels in (g), (h), (i), (j), (k), and (l) are the intensities of certain representative eigenfunctions in Fig.~\ref{Figure-1}. The right panels in (g), (h), (i), (j), (k), and (l) are Husimi distributions corresponding to each intensity of eigenfunctions in the left panel. The line $p_{c}$'s (black thick lines) in (g) and (j) are the centers of the striped $p$ distributions with the values $p_{c}\simeq0.67$ and $0.82$, respectively. The classical phase spaces (Poincar\'{e}-surface of section) are displayed  that are superimposed with Husimi distributions at right panels.}
\label{Figure-4}
\end{figure*}

 The figures on the left panels in Figs.~3 and 4 are the intensities of a few of the representative eigenfunctions as indicated in Fig.~\ref{Figure-1}. In the case of the Landau--Zener interactions (Fig.~\ref{Figure-3}), the eigenfunctions are mixed around the center of the avoided crossing and exchanged across the avoided crossing~\cite{KS18}. In contrast, the panels on the left in Fig.~\ref{Figure-4} for the Demkov interaction show a highly disparate aspect. The whispering galley-type modes (WGMs) with ($l=2,m=22$) and ($l=1,m=26$) at $\epsilon=0$ (\textit{i.e.,} Fig.~\ref{Figure-4}(g) and Fig.~\ref{Figure-4}(j)) are deformed to period-$4$ orbits, diamond (i) and rectangular (l)-shaped orbits, respectively. This is the result of the Fermi resonance~\cite{YH15}. That is, the quantum number difference equals the period of the orbits: $|\Delta N_{l}|,|\Delta N_{m}|=(1,4)$. Here, $N_{l}$ and $N_{m}$ are the radial and angular quantum numbers, respectively.

\subsection{Poincar\'{e}-Birkhoff coordinates and Husimi distributions}
The classical billiard dynamic is completely described by using the bounce map defined by the Poincar\'{e}-Birkhoff coordinates $(s,p)$. Here, $s$ is a boundary length of the cavity and its conjugate momentum $p=\textrm{sin}(\chi)$ and $\chi$ is the incident angle with respect to the normal to the boundary wall~\cite{MM76,M89}. The Fig.~\ref{Figure-2} shows an example of Poincar\'{e}-surface of section (classical phase space) defined in terms of Poincar\'{e}-Birkhoff coordinates at $\varepsilon=0.05$. In addition, the $(s,p)$ fulfill the Poisson bracket relation: $\{s,p\}=1$. Thus, to study quantum billiard dynamics, it is natural to employ the Husimi distributions~\cite{K40,BG93} in the Poincar\'{e}-Birkhoff coordinate defined below:

\begin{align}
H(s,p)=\frac{k}{2\pi}\left|\frac{i}{kF}h'(s,p)\right|^2
\label{husi-1}
\end{align}
with the angular momentum-dependent weighting factor $F=\sqrt{n\cos(\chi)}$ and $h'(s,p)=\int d s'\partial\psi(s')\xi(s';s,p)$~\cite{AS04}.
Here, $\partial\psi(s')$ is the normal derivative of the boundary wavefunction, and $\xi(s';s,p)$ is the minimal-uncertainty wave packet (coherent state) centered at a position $(s,p)$ in the phase space. The minimum-uncertainty wave packet is given by
\begin{align}
\xi(s';s,p)=&\sum_{m\in\mathbb{Z}}\frac{1}{\sqrt{\sigma\sqrt{\pi}}}\exp\big[-\frac{1}{2\sigma^{2}}(s'-s)^{2}\nonumber\\
&-ikp(s'+Lm)\big].
\end{align}
Here, $L$ and $k$ are the total length of the boundary and wave number, respectively. We set the aspect ratio factor as $\sigma=\frac{1}{\sqrt{(\sqrt{2}/k)}}$ on this coherent state.

The right panels in Fig.~\ref{Figure-3}, denoted as (a), (b), (c), (d), (e), and (f), are the Husimi distributions corresponding to each eigenfunction in the left panels for the Landau--Zerner interactions. The Husimi distributions displayed in Fig.~\ref{Figure-3}(a) and (d) are exchanged with each other across the center of the avoided crossing. Meanwhile, those displayed in Fig.~\ref{Figure-3}(b) and 2(e) at the center of the avoided crossing are more intricate than the others (Figs.~\ref{Figure-3}(a), 2(d), 2(c), and 2(f)).

On the other hand, the right panels (g), (h), (i), (j), (k), and (l) in Fig.~\ref{Figure-4} show a highly distinct tendency. We can easily observe that the WGM modes ((g), (j)) are transformed to the period-$4$ orbits ((i), (l)). The bright 4 spots of Husimi distributions in Fig.~\ref{Figure-4}(i) and (l) support this observation. Moreover, we also observe that (h) and (k) at the center of the avoided crossing are transient modes that settle down into a diamond-shaped stable period-$4$ orbit as well as a rectangle-shaped unstable period-$4$ orbit corresponding to a scar mode, respectively. We quantitatively analyze these features within the framework of information theory in Sec.~IV.

\section{Overlap analysis}\label{Helm}
\begin{figure}
\centering
\includegraphics[width=8.3cm]{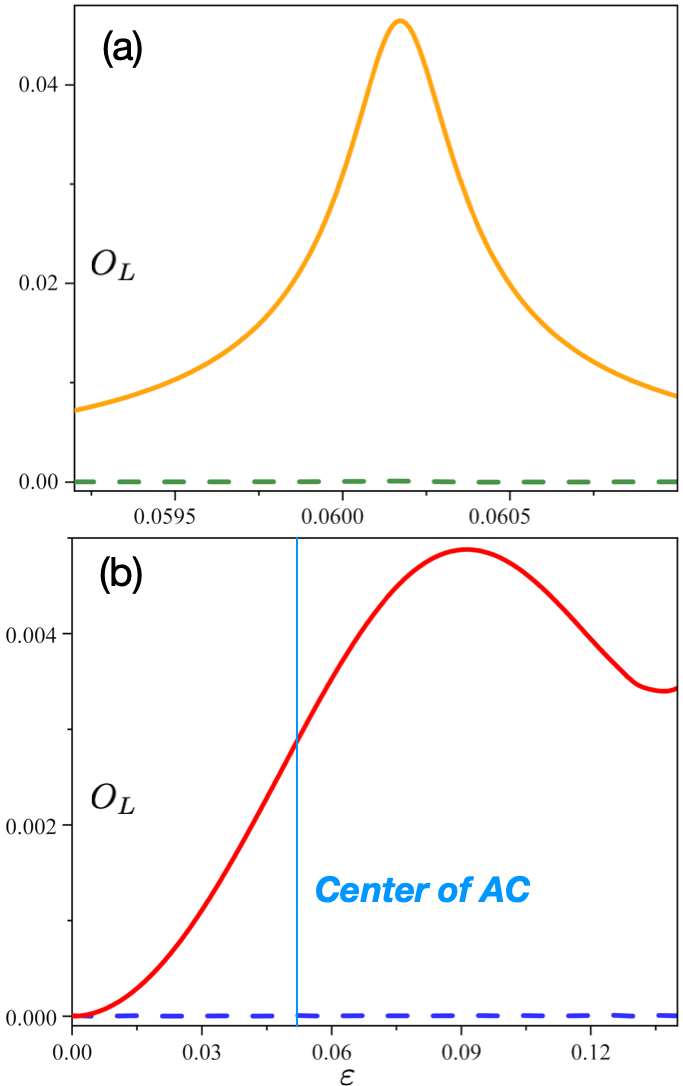}
\caption {Overlaps of wavefunction and that of Husimi distributions. (a) Overlap for the Landau--Zener interaction: The green dashed line at the bottom is the overlap of the wavefunctions, whereas the orange solid line is that of the Husimi distributions. It is maximized at approximately $\epsilon\simeq0.0603$. (b) Overlap for the Demkov interaction: The blue dashed line at the bottom is the overlap of the wavefunctions, whereas the red solid line is that of the Husimi distributions. It is maximized at approximately $\epsilon\simeq0.09$. The thin vertical line indicates the center of the avoided crossing.}
\label{Figure-5}
\end{figure}

Let us first consider overlap analysis before addressing informational analysis in detail. An overlap of eigenfunctions with respect to different eigenvalues even in chaotic system always  yields zero. Hence, it cannot provide insight into the interactions. In contrast, an overlap of Husimi distributions is generally not zero because these are not eigenvectors or eigenfunctions of the Hermitian operator~\cite{J88}. An overlap of Husimi distributions indicates the degree of occupation in the same region in a phase space~\cite{AC03}. Husimi distributions occupy a larger space of the same region in a phase space as their overlap increases.

For this analysis, we define an overlap of eigenfunctions $O_{L}^{\psi}(n,m)$ and that of Husimi distributions $O_{L}^{\tilde{H}}(n,m)$ as follows:

\begin{align}
O_{L}^{\psi}(n,m)=\frac{1}{X_{n}X_{m}}\int\int dxdy |\psi^{*}_{n}(x,y)\psi_{m}(x,y)|,
\end{align}
where $X_{n}=\sqrt{\int dxdy|\psi_{n}(x,y)|^{2}}$ is a normalization factor. Here, the integral is carried out the interior of the quadrupole billiard.

To establish the overlap of the Husimi distributions in a phase space, we transform the Husimi distribution ${H}(s,p)$ into $\tilde{H}(s,p)=\sqrt{\frac{k}{2\pi}}\frac{i}{kF}h'(s,p)$. This is appropriate for the inner product, i.e., $\tilde{H}(s,p)\tilde{H}^*(s,p)={H}(s,p)$.  Then, we define an overlap of the Husimi distributions as
\begin{align}
O_{L}^{\tilde{H}}(n,m)=\frac{1}{Q_{n}Q_{m}}\int\int dsdp |\tilde{H}^{*}_{n}(s,p)\tilde{H}_{m}(s,p)|,
\end{align}
where $Q_{n}=\sqrt{\int dsdp|\tilde{H}_{n}(s,p)|^{2}}$ is a normalization factor.

\subsection{Overlap on the Landau--Zener interaction}
 The green dashed line at the bottom of Fig.~\ref{Figure-5}(a) is the overlap of eigenfunctions $O_{L}^{\psi}(n,m)$, whereas the orange solid line in the Fig.~\ref{Figure-5}(a) is the overlap of the Husimi distributions $O_{L}^{\tilde{H}}(n,m)$. Both of these are for the Landau--Zener interaction in Fig.~\ref{Figure-3}. We observe that the value of the green dashed line at the bottom is zero throughout the interaction. This verifies the orthogonality in the Hermitian system. In contrast, the overlap of the Husimi distributions shows a distinct behavior. That is, it is maximized at the center of the avoided crossing ($\epsilon=0.0603$). This indicates that the degree of occupation in the same region in a phase space is maximized at the center of the interaction.

\subsection{Overlap on the Demkov interaction}
In Fig.~\ref{Figure-5}(b), the blue dashed line at the bottom is the overlap of wavefunctions $O_{L}^{\psi}(n,m)$, whereas the red solid line is that of the Husimi distributions $O_{L}^{\tilde{H}}(n,m)$. We can easily observe that the orthogonality of eigenfunctions is still valid on the Demkov interaction because the value of the blue dashed line at the bottom of Fig.~\ref{Figure-5}(b) is also zero throughout the interaction. However, the overlap of the Husimi distributions $O_{L}^{\tilde{H}}(n,m)$ exhibits a behavior different from that of the orange solid line in Fig.~\ref{Figure-5}(a). That is, the center of the avoided crossing is located at $\epsilon\simeq0.055$, whereas the maximum value of $O_{L}^{\tilde{H}}(n,m)$ is located at $\epsilon\simeq0.09$. These two values are significantly different from each other. We analyze these features by exploring Shannon entropy in the next section.

\section{Shannon information entropy in a phase space} \label{GHeff}
Shannon information entropy can be interpreted as the uncertainty of a probability distribution or a measure of the average information content. It was first proposed by Claude Shannon in his communication theory~~\cite{C48}. However, it has been widely utilized in various fields recently. Shannon entropy has been employed for discrete multidimensional hydrogenic states~\cite{ID20}, helium atom in screened Coulomb potentials~\cite{C21} in chemistry, personalized medicine~\cite{AJ10}, Alzheimer’s disease~\cite{AN19} in biology, economics~\cite{O19}, and ecology~\cite{WN19}.

In our recent study, we utilized the Shannon entropy in the configuration space and investigated its relationship with the localization properties~\cite{KJ21}. In this study, we extend our investigation to entropies in a phase space with Husimi distributions. The Shannon entropy in a phase space can be defined as
\begin{align}
\emph{\textsf{S}}= -\int\int dsdp H(s,p)\ln H(s,p)
\end{align}
with the normalization condition $\int dsdp H(s,p)=1$. In this case, one random variable is the arc length of the cavity $S$, and the other random variable is its conjugate momentum $P$ with their joint probability $\rho(s,p)=\textsf{P}(S=s,P=p)$, resulting in $H(s,p)=\rho(s,p)$. To perform the numerical calculations, we introduce the discrete probability distributions: we discretize the phase space $(s,p)$ into $100\times100$ pieces that play the role of the $10000$-mesh points at each $\epsilon$. Then, we assign the probability density $\rho(s_{i},p_{j})$ to each mesh point $(s_{i},p_{j})$ under the normalization condition $\sum_{i,j}\rho(s_j,p_j)=1$ by interpreting the $10000$-mesh points as the discrete phase space-coordinate. We utilize this definition of the Shannon entropy to analyze the two types of interactions in a phase space. These are compared with the overlap of Husimi distributions below.

\begin{figure}
\centering
\includegraphics[width=8.5cm]{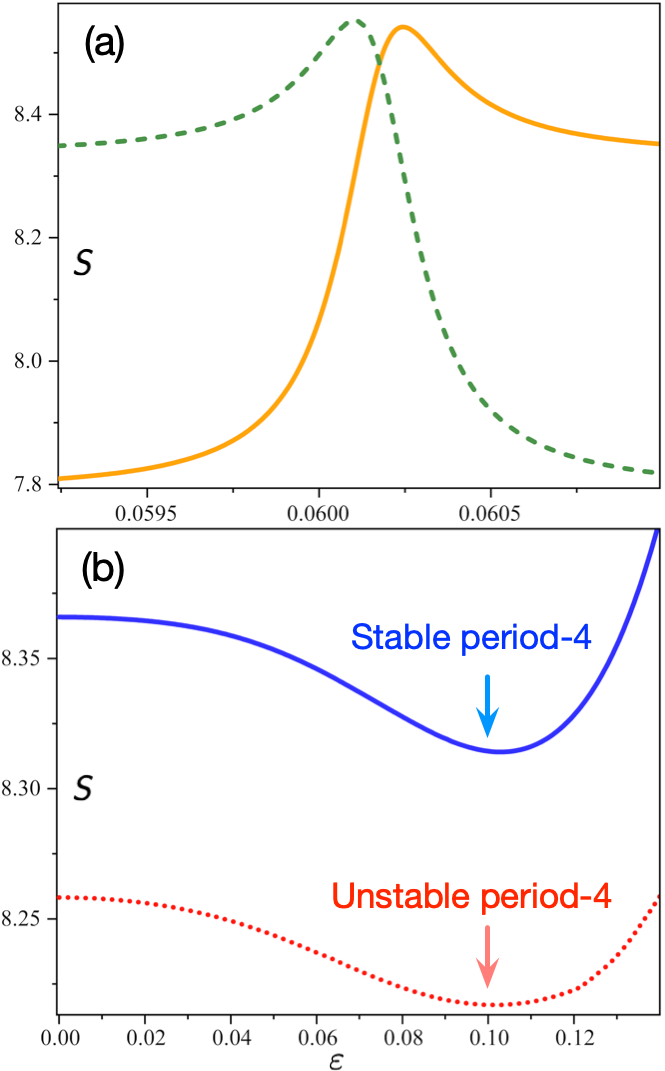}
\caption {Shannon entropies in a phase space. (a) Shannon entropies of two eigenfunctions in a phase space for the Landau--Zener interaction. These are maximized around the center of the avoided crossing and exchanged across the avoided crossing. (b) Shannon entropies of two eigenfunctions in a phase space for the Demkov interaction. These are minimized at $\epsilon\approx 0.1$. The values of the Shannon entropy of the stable period-$4$ orbit are larger than those of the unstable period-$4$ orbit.}
\label{Figure-6}
\end{figure}

\subsection{Shannon entropy of the Landau--Zener interaction}
The Shannon entropies of the two eigenfunctions in a phase space for the Landau--Zener interaction are plotted in Fig.~\ref{Figure-6}(a). These are maximized around the center of the avoided crossing ($\epsilon\simeq0.0603$) and exchanged across the avoided crossing. Note that these two Shannon entropies have similar maximum values, \textit{i.e., } $\emph{\textsf{S}}_\tn{max}\simeq8.57$ for the green dashed line and $\emph{\textsf{S}}_\tn{max}\simeq8.56$ for the yellow solid line. These features are in good agreement with the results of the Shannon entropies in a configuration space~\cite{KS18}. Moreover, our previous study verified that the Shannon entropy as well as inverse participation ratio and image contrast can measure the localization properties of eigenfunctions on configuration space~\cite{KJ21}. Thus, the maximum values of Shannon entropies around the center of the avoided crossing can directly indicate that the Husimi distributions are maximally delocalized in the phase space around the center of the avoided crossing.

 Therefore, we can explain why we obtain the maximum value of the overlap of the Husimi distributions $O_{L}^{\tilde{H}}(n,m)$ at the center of avoided crossing. The increased delocalization or diffusion of the two Husimi distributions causes these two distributions to occupy a larger space of the same region in a phase space. An examination of the patterns inside the two boxes in Figs.~3(b) and 3(e) reveals this fact. In this manner, the overlap of Husimi distributions can capture the effects of the avoided crossing, even though the overlap of the eigenfunction cannot do so.

\subsection{Shannon entropy of the Demkov interaction}
Figure 6(b) displays the Shannon entropies of the two eigenfunctions in a phase space for Demkov interaction. The behaviors of these Shannon entropies differ significantly from those in Fig.~\ref{Figure-5}(a). They decrease (rather than be maximized) as they go through the center of avoided crossing with increased deformation. The Husimi distributions of the eigenfunctions at $\epsilon=0.0$ (circles) have rotational symmetries, so that these distributions spread along the boundary. However, once the deformation increases, the Husimi distributions start to move toward the periodic orbits owing to the Fermi resonance~\cite{YH15}. This process continues until they settle down into periodic orbits. This can be revealed by the local minimum values of Shannon entropies at $\epsilon\simeq0.1$ because the Shannon entropy can also measure the degree of (de)localization~\cite{KJ21}. Here, we can quantitatively state the main difference between the Landau--Zener interaction and Demkov interaction. The former induces delocalization in a phase space, whereas the latter induces localization in a phase space.

Moreover, the values of the blue solid curve are larger than those of the red dotted curve throughout the process, not being exchanged with the red dotted-curve. In other words, the values of the Shannon entropy of the stable period $4$ are larger than those of the unstable period $4$ (scar). This is because the stable period-$4$ orbit settles down on the island structures having finite areas, whereas the unstable period-$4$ orbit corresponding to scar settles down on the points of measure zero. Therefore, we can presume that the Shannon entropy in a phase space can be a new indicator of scar formation.

Finally, these behaviors explain why the maximum values of $O_{L}(\tilde{H}_{(n,m)})$ are obtained in the transient regime. Two Husimi distributions of eigenfunctions (WGM) in the circle ($\epsilon=0$) are placed at highly distinct values of the $p=\textrm{sin}(\chi)$ (center of $p$: $p_{c}\simeq0.67$ and $p_{c}\simeq0.82$, respectively) as shown in Figs.~4(g) and (j) such that these two WGMs occupy distinct striped regions in the phase space. However, when the Demkov interaction occurs, these two Husimi distributions shift toward the period-$4$ orbits. These shifts lead to the occupation of a larger space of the same region in a phase space. It should be noted that the maximum value of $O_{L}(\tilde{H}_{(n,m)})$ is attained in the transient regime at $\epsilon=0.09$, rather than at $\epsilon=0.1$, where the two Husimi distributions are maximally localized on the stable period-$4$ orbit and unstable period-$4$ orbit, respectively. In this manner, the phase space overlap and Shannon entropy in a phase space can be related to each other under the influences of avoided crossing.

In addition, it should be noticed that RAT partly can explain the results above. The value of $O_{L}\simeq 0.0$ at $\epsilon=0.0$ comes from the condition that one unperturbed mode (with $p_{c}\simeq 0.82$) must be placed above the $t=4:r=1$ resonance chain with $P_{{t=4}:{r=1}}=\cos(\pi r/t)=1/\sqrt{2}\simeq 0.707$ while the other one (with $p_{c}\simeq 0.67$) must be placed below the $P_{{t=4}:{r=1}}$ so that RAT can occur. Here, the $P_{{t=4}:{r=1}}$ corresponds to the adiabatically invariant curve interpolating through the nonlinear resonance~\cite{FR19}. The increased value of the $O_{L}$ can be due to the increased coupling strength by increased area of separatrix of the resonance chain. However, the center of avoided crossing is located at $\epsilon\simeq0.055$ far from $\epsilon=0.09$ (maximum of value of the $O_{L}$) or $\epsilon=0.1$  (minimum of value of the $\emph{\textsf{S}}$). It is also worth mentioning that the Demkov interaction does not always induce a scar formation. For example, regarding an excited-state quantum phase transition, the Husimi distribution is more localized at the center of avoided crossing with a transition between two regular states, i.e., tori of rotational motion and that of the libration motion, without giving rise to any scar state~\cite{IE21}. Those problem should be addressed in the future works.

Lastly, the absolute value of the overlap of Husimi distributions $O_{L}^{\tilde{H}}(n,m)$ for the Demkov interaction is ten times smaller than that of $O_{L}^{\tilde{H}}(n,m)$ for the Landau--Zener interaction. This is because there is negligible resemblance in the Demkov cases, whereas we can observe resemblance in the Landau--Zener case shown in the two boxes of Fig.~\ref{Figure-3}(b) and 3(e).

\section{Fisher information in a phase space}\label{QNM}
Fisher information is a mathematical measure of the sensitivity of an observable data to variations in its underlying parameters~\cite{BM13,DS21}.
Formally, the partial derivative with respect to the system parameter ($\epsilon$) of the natural logarithm of the likelihood function is called the score. Then, the variance of the score is the Fisher information, which is defined as
\begin{equation}
\emph{I}(\epsilon)=\sum_{i,j=1}^{100}\left(\frac{\partial}{\partial\epsilon}\ln\rho(s_j,p_j;\epsilon)\right)^{2}\rho(s_j,p_j;\epsilon),
\end{equation}
where $\epsilon$ denotes the system parameter and  $\rho(s_j,p_j;\epsilon)$ is a probability density function.
In this section, we explore the Fisher information to quantitatively distinguish between the narrow and broad interactions in the deformation parameter $\epsilon$.

\begin{figure}
\centering
\includegraphics [width=8.5cm]{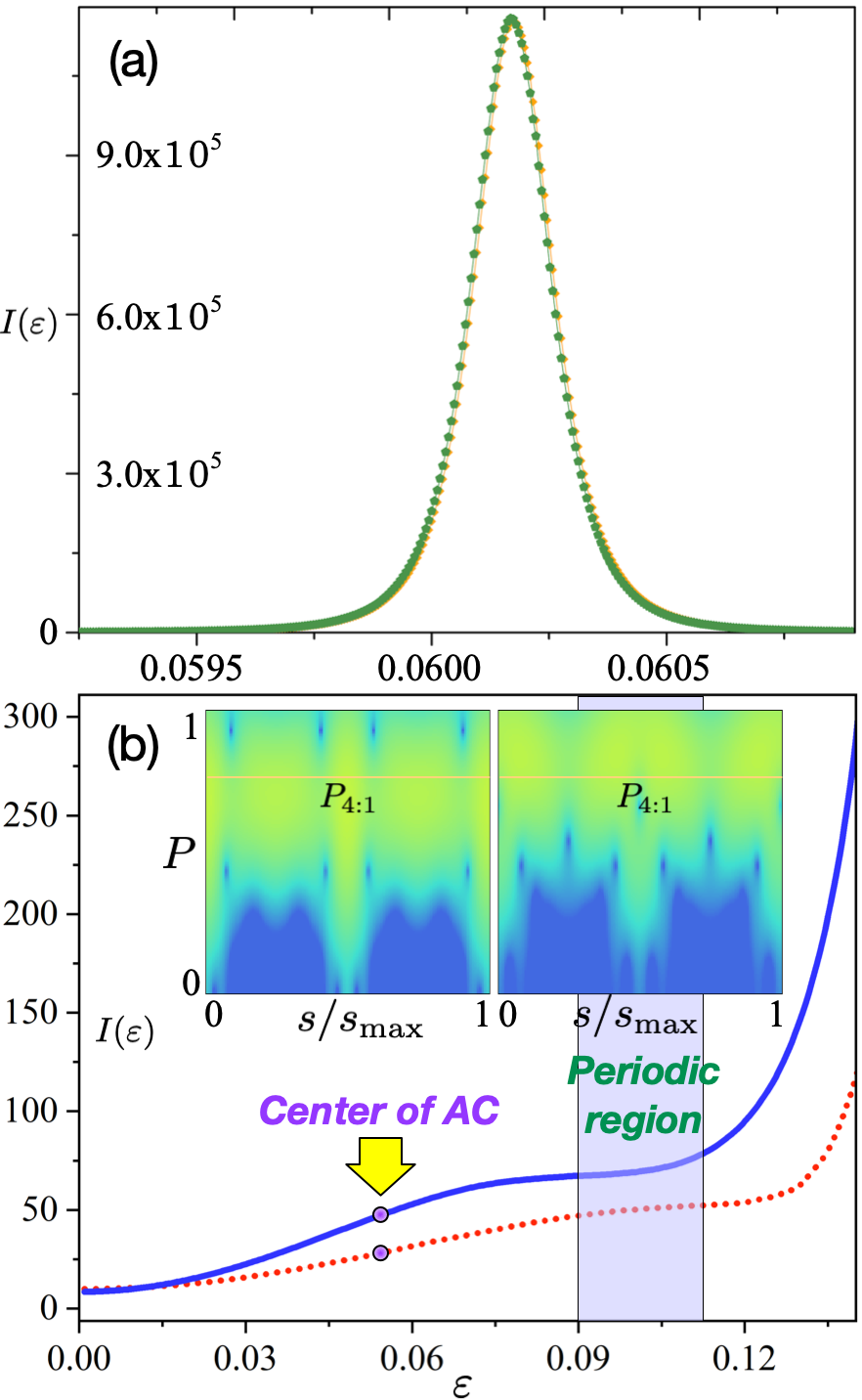}
\caption {Fisher information in a phase space. (a) Fisher information of the Landau--Zener interactions in a phase space. These increase dramatically as the center of the avoided crossing is approached. (b) Fisher information of the Demkov interactions in a phase space. The insets are log plots of Husimi distributions at $\epsilon=0.125$.}
\label{Figure-7}
\end{figure}

\subsection{Fisher information of the Landau--Zener interaction}
Figure 7(a) shows the Fisher information $I(\epsilon$) as a function of the deformation parameter $\epsilon$ in the case of the Landau--Zener interactions. $I(\epsilon$) has relatively high marginal values (of the order of $10^{0}$) well away from the avoided crossings ($\epsilon\simeq0.059, \epsilon\simeq0.061$). However, it increases dramatically as the center of interaction ($\epsilon\simeq0.0603$) is approached and attains values of the order of $10^{5}$. This directly indicates that the Landau--Zener interaction results in an extensive variation in the eigenfunctions over a highly narrow parameter range. Consequently, we can quantify the degree of the effects of the avoided crossing on the eigenfunctions depending on the system parameter. Furthermore, note that $I(\epsilon$) of the two interacting modes shows no significant exchange across the avoided crossing, unlike the Shannon entropies in Fig.~\ref{Figure-6}. This implies that the degree of variations in eigenfunctions under an interaction on the system parameter is similar, whereas the degree of eigenfunctions localization under an interaction on the system parameter is not.

\subsection{Fisher information of the Demkov interaction}
Contrary to the Landau--Zener interaction, the Fisher information of the Demkov interactions exhibits various features as shown in Fig.~\ref{Figure-7}(b). First, the absolute values of $I(\epsilon$) on the Demkov interactions (which are of the order of $10^{1}\sim10^{2}$) are extremely small compared to that on the Landau--Zener interactions. Such small values quantitatively support that the Demkov interaction occurs over a broad region, whereas the Landau--Zener interactions occur over a highly narrow region.

 What is more noteworthy is that the Fisher information $I(\epsilon$) on Demkov interaction gradually (almost linearly) increases across the avoided crossing rather than being maximized at the center of the avoided crossing ($\epsilon=0.055$). This implies that the rate of variations in the eigenfunctions increase across the center of the avoided crossing. Subsequently, the Fisher informations are barely changed in the range $0.09\leq\epsilon\leq 0.11$. This behavior can be attributed to their settling down into the periodic orbits. This is consistent with the results shown in Fig.~\ref{Figure-6}(b), \textit{i.e.,} the Shannon entropies on the Demkov interactions are locally minimized around the $\epsilon\simeq 0.1$. The substantial increase in $I(\epsilon$) in the region $\epsilon\geq0.12$ is owing to leakage from the periodic orbits shown in the insets of Fig.~\ref{Figure-7}(b). We can check that there are considerable amount of probability below the $P_{4:1}\simeq 0.707$ at $\epsilon=0.125$.

\section{Conclusions}
We compare the Landau--Zener interaction with the Demkov interaction in a quadrupole billiard from the perspective of information theory. The Shannon entropies in a phase space for the Landau--Zener interaction increase as the deformation parameter approaches the center of the avoided crossing. This indicates that Husimi distributions become more delocalized in the phase space. This results in larger values of the overlap of the two Husimi distributions, \textit{i.e.,} the two Husimi distributions occupy a larger space of the same region on a phase space.

Meanwhile, the Shannon entropies in a phase space for the Demkov interaction decrease as they go through the center of avoided crossing with increasing deformation. They achieve local minimum values as they settle down into periodic orbits. Furthermore, the Shannon entropy of an unstable periodic orbit is smaller than that of a stable periodic orbit throughout the course of interaction. These results can present a new method for measuring scar formation. The overlap of the two Husimi distributions is maximized before the Husimi distributions maximally settle down into a periodic orbit.

Lastly, the Fisher information of the Landau--Zener interaction is $10^{3}\sim10^{4}$ times larger than that of the Demkov interaction. This observation quantitatively verifies that the Landau--Zener interaction occurs over a highly narrow region compared with the Demkov interaction.

It is also worth remarking the relations between present works and quantum information processing. Understanding the quantum entanglement is essential to physical realization in the field of quantum information science because the quantum entanglement is a key resource of the implementing quantum computation and other quantum technology. In this reason, the relations between quantum chaos and avoided crossing with respect to the quantum entanglement have been steadily investigated in other research field such as multi-qubit system~\cite{XS04}, ising model~\cite{JA07}, and quantum tomograms~\cite{BS20} so far. Accordingly, our results can be helpful to exploit the quantum entanglement in the context of avoided crossings and quantum chaos.

\begin{acknowledgments}
This work was supported by the Korea Research Foundation (Grant No.~2020R1A2C3009299), the Ministry of Science and ICT of Korea under ITRC program (Grand No. IITP-2021-2018-0-01402) and Ministry of Education (Grant No.~NRF-2021R1I1A1A01052331). K.J. acknowledges the support from the National Research Foundation of Korea, a grant funded by the Ministry of Science and ICT (Grant No.~NRF-2020M3E4A1077861) and the Ministry of Education (Grant No.~NRF-2021R1I1A1A01042199).
\end{acknowledgments}

\end{document}